\documentclass[10pt, conference, compsocconf]{IEEEtran}
\IEEEoverridecommandlockouts
\usepackage{cite}
\usepackage{amsmath,amssymb,amsfonts}
\usepackage{algorithmic}
\usepackage{graphicx}
\usepackage{textcomp}
\usepackage{xcolor}
\usepackage{comment}
\usepackage{multirow}
\usepackage{multicol}
\usepackage{xspace}
\usepackage{makecell}
\usepackage{balance}
\usepackage{url}
\usepackage{hyperref}
\newcommand{\tool}{\textsc{ReplIoT}\xspace}

\newcommand\new[1]{#1}

\def\BibTeX{{\rm B\kern-.05em{\sc i\kern-.025em b}\kern-.08em
    T\kern-.1667em\lower.7ex\hbox{E}\kern-.125emX}}

\usepackage{todonotes}

\makeatletter
\newcommand{\linebreakand}{%
  \end{@IEEEauthorhalign}
  \hfill\mbox{}\par
  \mbox{}\hfill\begin{@IEEEauthorhalign}
}
\makeatother

\newcommand{\cmark}{\color{green} \checkmark}
\newcommand{\xmark}{\color{red} X}    

\usepackage[symbol]{footmisc}

\usepackage[showboxes]{textpos}

\usepackage{textcase}

\begin{document}

\title{Is Your Kettle Smarter Than a Hacker?\\ A Scalable Tool for Assessing Replay Attack Vulnerabilities on Consumer IoT Devices}



  \author{\IEEEauthorblockN{Sara Lazzaro, Vincenzo De Angelis}
 \IEEEauthorblockA{\textit{Mediterranea University of Reggio Calabria} \\
 \textit{University of Calabria}}
 
 \and
 \IEEEauthorblockN{Anna Maria Mandalari}
 \IEEEauthorblockA{\textit{University College London} \\
 \textit{}}
 \and
 \IEEEauthorblockN{Francesco Buccafurri}
\IEEEauthorblockA{\textit{Mediterranea University of Reggio Calabria} \\
 \textit{}}
 \thanks{
 
\textit{The two lead authors contributed equally to this work.} 


}

}

\maketitle

\begin{abstract}

Consumer Internet of Things (IoT) devices often leverage the local network to communicate with the corresponding companion app or other devices. This has benefits in terms of efficiency since it offloads the cloud. ENISA and NIST security guidelines underscore the importance of enabling default local communication for safety and reliability. Indeed, an IoT device should continue to function in case the cloud connection is not available.
While the security of cloud-device connections is typically strengthened through the usage of standard protocols, local connectivity security is frequently overlooked.
Neglecting the security of local communication opens doors to various threats, including replay attacks.
In this paper, we investigate this class of attacks by designing a systematic methodology for automatically testing IoT devices vulnerability to replay attacks. Specifically, we propose a tool, named \tool, able to test whether a replay attack is successful or not, without prior knowledge of the target devices. We perform thousands of automated experiments using popular commercial devices spanning various vendors and categories. Notably, our study reveals that among these devices, 51\% of them do not support local connectivity, thus they are not compliant with the reliability and safety requirements of the ENISA/NIST guidelines.
We find that 75\% of the remaining devices are vulnerable to replay attacks with \tool having a detection accuracy of 0.98-1. Finally, we investigate the possible causes of this vulnerability, discussing possible mitigation strategies.
\end{abstract}


\begin{textblock}{13}(0,-10)
  \noindent
  \footnotesize
  \centering
  \textblockcolour{green}
  \vspace{3mm}
  
\textbf{If you cite this paper, please use the PERCOM reference: S. Lazzaro, V. De Angelis, A. M. Mandalari and F. Buccafurri, "Is Your Kettle Smarter Than a Hacker? A Scalable Tool for Assessing Replay Attack Vulnerabilities on Consumer IoT Devices," 2024 IEEE International Conference on Pervasive Computing and Communications (PerCom), Biarritz, France, 2024, pp. 114-124, doi: 10.1109/PerCom59722.2024.10494466.}

\vspace{1mm}
\end{textblock}

\section{Introduction}

Consumer Internet of Things (IoT) devices, such as smart
TVs, speakers, surveillance cameras, and appliances, offer numerous advantages to their users~\cite{atzori2010internet, stojkoska2017review, aldossari2020consumer}. 
These devices can be managed remotely by the users through smartphone apps connected to cloud platforms. 
However, in most cases, to optimize the performance, when the smartphone is in the same network of the device, local communication between the app and the device is enabled.
In this way, the latency time is reduced~\cite{latencyTimes} and the device still works even if the remote connection is not available.  
This latter point is crucial for reliability and safety reasons.
This principle is reported in the ENISA security guideline GP-TM-17~\cite{ENISA-IoT}, which states that IoT devices should continue to work if the cloud back-end fails. 
Indeed, despite a user may be in the same network of the IoT device, physical access to the latter may still be hard (e.g., for impaired users or in case the device is difficult to reach because of its position). 
In these cases, the loss of connectivity to the Internet should not limit the possibility for the user to control a device that may have a safety impact (e.g., to turn off a smart oven, open a smart lock, start a camera recording, etc.). 

On the other hand, enabling local connectivity requires proper security measures as highlighted by the security guidelines of ENISA~\cite{ENISA-IoT} and NIST~\cite{nist}.
As a matter of fact, each device should not rely on the security of firewalls or other security practices implemented in the access point but should have its own security mechanisms.
\new{As witnessed recently by Girish et al.~\cite{10.1145/3618257.3624830}, despite its importance, the security of IoT local connectivity has not been deepened in the scientific literature.}

In this paper, we focus on one of the main threats arising in the local network, i.e., replay attacks~\cite{syverson1994taxonomy,singh2020countermeasures}.
Replay attacks are performed by eavesdropping the network traffic, intercepting it, and replaying the packets after a given time.

In smart homes~\cite{7917634}, this class of attacks is very relevant for two reasons.
First, replay attacks can be easily performed by a single malicious device inside the local network~\cite{shaikh2018internet,sinanovic2017analysis, hafeez2020iot}. We expect this scenario to become increasingly likely given the proliferation of IoT devices~\cite{proliferation,ren2019information}.
Second, replay attacks may be part of a more complex attack strategy, serving as a means to furnish adversaries on insights into the controllability of a device~\cite{securityProblems}.
While the prior knowledge of the target device may enable ad hoc solutions to perform replay attacks~\cite{yoon2020new}, a challenging question is whether they can be automatically performed on generic devices. 
\new{To the best of our knowledge, no papers in the scientific literature pursue this ambitious goal.}

In this work, we propose \tool, a tool for automatic testing replay attacks in a smart home environment. The tool is designed to be \textit{agnostic} and works with any type of device and communication protocol. 
\new{Our innovative methodology is based on black-box processing of the sniffed traffic to increase the chances of making the attack successful.}
Being our tool device-agnostic, it can also be used in a household environment by non-technical users~\cite{Mobicom2023Certification}. To this aim, we equip \tool with a detection module that works on the device's responses to automatically detect whether the attack is successful or not. 
This module is particularly relevant in the case in which the user launching the tool has no physical access to the device to observe the effect of a replayed command. Furthermore, there exist commands that may not trigger an external change in the device (e.g., a change in the sensitivity level of a motion detection sensor). 

Our key research contributions include:
\begin {itemize}
\item We develop an automated methodology for large-scale testing replay attack vulnerabilities on IoT devices;

\item We demonstrate the feasibility of detecting the success of the attack;

\item We assess the (in)effectiveness of $41$ popular IoT devices in preventing such attacks;

\item and finally, we examine the potential causes behind these vulnerabilities, shedding light on the factors contributing to these security weaknesses.
\end{itemize}

We find that 21 IoT devices do not support local connectivity in contrast to the reliability and safety guidelines of ENISA~\cite{ENISA-IoT}.
Out of the 20 remaining devices, 15 are vulnerable to replay attacks through our tool. We perform thousands of automated experiments to validate \tool and show the effectiveness of the designed detection module. \new{The code of our tool and the data collected in our experiments are publicly available at: \url{https://github.com/SafeNetIoT/ReplayAttack}}. 


\textbf{Responsible Disclosure.} 
\new{We responsibly disclosed our results with the IoT manufacturers in this study. We
received responses from one manufacturer. We include these responses (with
permission) in Section~\ref{sec:discussion}}.


\section{Assumption, Goals and Non-Goals} \label{sec:assumptions}
In this section, we set the assumptions, the goals, and the non-goals for this work.

\subsection{Threat Model}
We consider the following threat model.

\noindent \textbf{Victim.}  The victim is any person who uses or benefits from consumer IoT devices.

\noindent \textbf{Adversary.} The adversary is any party that can access the local IoT device traffic. Examples include internal privacy and
security threats, and malicious IoT devices, placed within the local network, with the ability to sniff promiscuously~\cite{miettinen2017iot}. Remote attackers can also exploit router vulnerabilities (e.g., default passwords) to access the local network~\cite{antonakakis2017understanding}.
\new{We observe that when our tool is adopted for defensive purposes, it assumes the role of an adversary (without leading to threats). It can be placed directly on the access point or a device connected to the same network.}

\noindent \textbf{Threat.} The adversary can trigger commands on smart home IoT devices, thus instructing them to perform some actions without the victim's will.


\noindent \textbf{Plausibility.} Several security guidelines of ENISA~\cite{ENISA-IoT} 
and NIST~\cite{nist} (Logical Access to Interfaces) witness the importance of protecting the local network segment.
Furthermore, also the scientific literature acknowledges this problem~\cite{miettinen2017iot,capellupo2017security,serror2018towards}.
As highlighted by Miettinen et al.~\cite{miettinen2017iot}, also remote attacks through “NAT hole punching”~\cite{acar2018web}, may be effective by compromising a device (e.g., a smartphone) with access to the local network. 
\new{As a result, devices should not depend on the security provided by firewalls or other measures implemented in the access point. This aligns with the GP-TM-43 of ENISA~\cite{ENISA-IoT}.}

\subsection{Goals}
The main goal of this work is to design a methodology to automatically verify whether an IoT device might be vulnerable to replay attacks. 
In particular, this work
answers the following research questions (RQ):

\textbf{RQ1:} \textit{Can replay attacks against IoT devices be automated without prior knowledge of the device-specific features?}
To answer this question we build a tool (\tool) able to actively perform replay attacks against generic IoT devices. We also design a methodology to employ this tool for a large-scale study of replay attack vulnerabilities of IoT devices. 


\textbf{RQ2:} \textit{Is it possible to automatically detect whether a replay attack is successful?}
Since also non-technical users can adopt our tool in a real environment, we design a methodology to automatically detect whether a replay attack is successful on a given device. 
Overall, our tool is conceived to minimize human intervention.

\textbf{RQ3:} \textit{Does the attack apply to a variety of IoT devices?}
To answer this question, we employ commercial IoT devices spanning various vendors and categories.

\subsection{Non-Goals} \label{non-goals}
In this work, we do not consider the following
as goals.

\noindent \textbf{Development of an Intrusion Detection System} \cite{anthi2019supervised}. The goal of our tool is not to verify that a device is undergoing a replay attack. On the contrary, our tool acts preemptively by identifying potential device vulnerabilities, \new{regardless of whether these devices may be in a potentially protected network segment (see GP-TM-43 of ENISA~\cite{ENISA-IoT}).}


\noindent \textbf{Design of ad hoc procedures for vulnerability testing.} Our tool is based on network traffic analysis, thus it does not require any knowledge of the device under test. 


\noindent \new{\textbf{Usability testing.} We do not perform usability testing on our tool. Even though \tool is deployed to be used in the home environment and not just in a lab environment, some refinements are needed to enhance its user-friendliness. However, addressing these aspects is beyond the scope of this paper.}

\section{Methodology} \label{sec:methodology}
We answer our research questions by proposing a
tool to automatically test the replay attack vulnerabilities of an IoT device.
 Our tool works with consumer IoT devices that are managed through a proper companion app provided by vendors.
\tool is based on network traffic analysis. At a high level, it works as follows. First, we sniff the local traffic (if any) exchanged between the companion app and the target device.
Then, we replicate in a proper fashion the sniffed traffic and monitor the (possible) responses received by the device.
Finally, we analyze these responses to automatically check the success or failure of the performed attack.

\tool presents three modules: {\tt Training Module}, {\tt Attack Module}, {\tt Detection Module}.
Overall, each of the three modules corresponds to a different phase in which the tool works.
In the following sections, we describe the three modules in detail.

\subsection{Preliminary Considerations}

\subsubsection{Machine Learning Algorithms Selection}
\tool implements a detection module to automatically check the success or failure of the performed attack. This module includes a machine learning (ML) algorithm that falls in the class of anomaly detection, specifically novelty detection~\cite{PIMENTEL2014215}. We train the algorithm using the responses sent by the device to the companion app. When \tool performs the attack, it also detects its success or failure by feeding the ML algorithm with the obtained responses.

The intuition is that, when the replay attack works, the responses received by \tool are similar to those exchanged by the device with its companion app. 
We refer to this type of response as \textit{regular response}. 
On the other hand, when the attack does not work, \tool does not receive any response or the device sends an error message. 
Such error messages should appear different from legitimate responses (with which the ML algorithm is trained). We refer to the responses carrying the error messages as \textit{irregular responses}. 
The ML module distinguishes between regular and irregular responses. 
We observe that, in the ML terminology, an irregular response is considered a novelty.
Novelty detection requires that the training instances do not contain any anomalies. 
This fits our scenario in which no error message is expected during the legitimate use of the device with the companion app. 

\subsubsection{Observed Types of Responses} \label{sec:observed_responses}
From a preliminary study, we observe four types of responses in the local traffic. 
\begin{itemize}
    \item \textbf{Full Cleartext Responses}:  words from natural language, mostly organized in a structured format (e.g., JSON).
    \item \textbf{Standard Encrypted Responses}: responses encrypted through standard protocols, e.g., TLS, QUIC, etc.
    \item \textbf{Non-standard Encrypted Responses}: responses encrypted through non-standard protocols, which can also carry cleartext metadata along with encrypted data.
    \item \textbf{Encoded Responses}: responses that are neither encrypted nor expressed in natural language, but they contain fixed bytes encoding some messages of proprietary protocols.
\end{itemize}

When a device executes a command (triggered by \tool), if it replies with cleartext or encoded messages, we expect they present some similarity with the responses provided by the device when it is triggered by the legitimate companion app.
Indeed, when no encryption mechanism is adopted, the responses to the same command are exactly the same or present minor changes (e.g., identifiers or timestamps).
Then, our ML algorithm can detect them as regular responses.
On the other hand, when the devices do not execute a command (the replay attack does not work), we expect that the responses (if any) include error messages not included in the legitimate responses. 
Then, the ML algorithm can detect these responses as irregular.

Regarding standard encrypted responses, we expect that the replay attack does not work and the ML module is not applicable, due to the fact that the encryption of two identical cleartext messages results in different encrypted payloads.

Finally, when a device uses non-standard encrypted responses, we cannot predict a well-defined behavior.
The response may present insufficient randomness, suffer from key reuse, or contain some cleartext metadata. Then, it is possible that the responses to the same command present a given degree of similarity, and the replay attack may work. 
Otherwise, similarly to standard encrypted responses, we expect the replay attack does not work.

\begin{figure*}[t!]
	\centering
    \includegraphics[width=1\textwidth ]{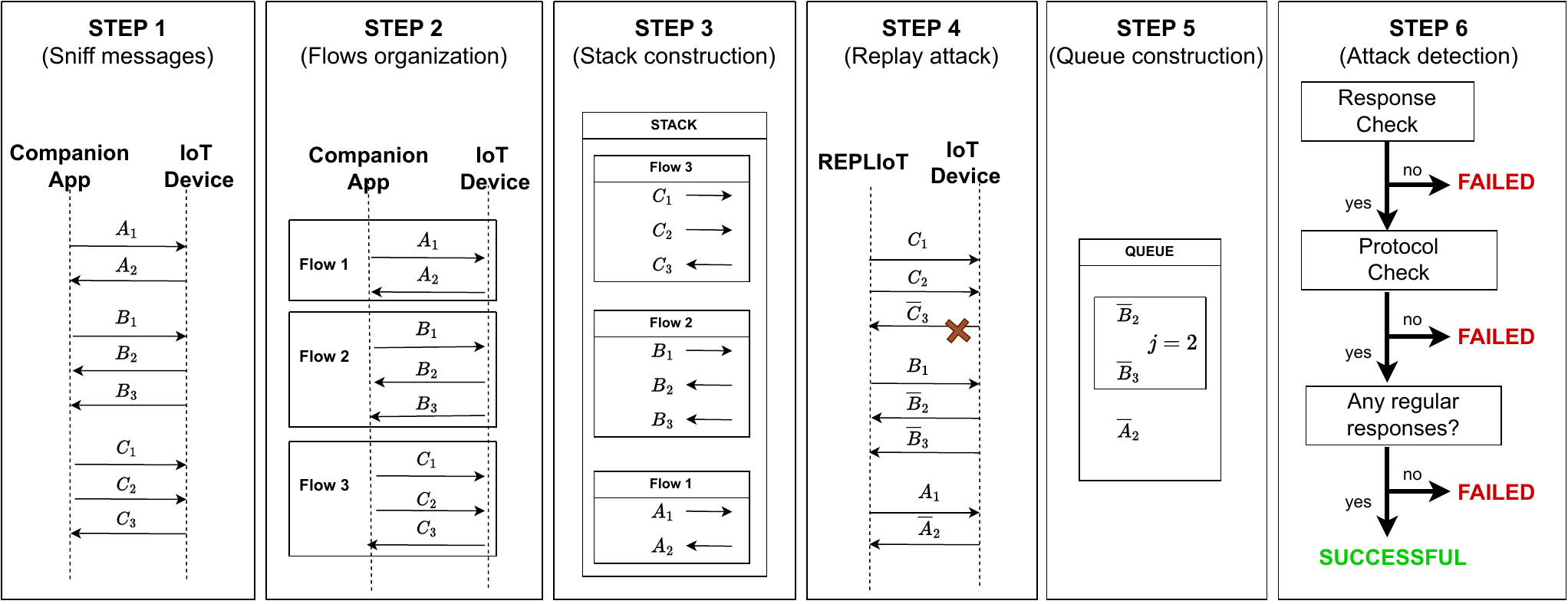}
     \caption{\tool Operation Steps. 
     }
  \label{fig:proposed_framework}
\end{figure*}

\subsection{Training Module} \label{sec:training_phase}

The training phase is devoted to the collection of some responses of the target device to train the ML model.
We obtain the responses by triggering some functions on the device through its companion app and sniffing the traffic. We collect all network traffic traversing the testbed using \emph{tcpdump}.

We set as a usability requirement the fact that the ML model should perform well with a few training data.

Once collecting them, \tool processes the responses, via {\tt Pyshark}~\cite{pyshark}, a Python tool adopting Wireshark dissectors. Specifically, we extract the payloads of the transport level (TCP or UDP) messages to train the ML model.



\subsection{Attack Module} \label{sec:attack_phase}
We trigger the attack when a function (e.g., switching on/off the light of a smart bulb, watching live from a camera, etc.) on the target device is performed. The aim is to understand if our tool can trigger the same function at a later time. 

The first step of this phase consists of sniffing the traffic exchanged between the legitimate companion app and the target device via {\tt tcpdump} when this function is triggered. 

We process the captured traffic, through {\tt Pyshark}, by extracting the transport level payloads. However, differently from the training phase, we collect both the requests and the responses.
Subsequently, we organize the extracted payloads in \textit{flows}.
Each flow is represented by a list of consecutive requests from the companion app to the device and a list of consecutive responses from the device to the companion app. 
Once the flows are collected, \new{the responses can be discarded, and} we are ready to launch the attack.

The attack consists of replaying the requests of each flow and storing the received responses (if any).
The list of flows is organized as a stack so that the last flow will be the first to be replayed.
The rationale behind this {empiric} design choice is that, for some devices, it may happen that, to trigger a specific function, several requests and responses (and then several flows) are exchanged.
However, the first flows may not contain the actual command (that triggers the function) but just accessory information (e.g., the information needed to exchange some secret or synchronization messages). On the other hand, it is more likely that the actual command is contained in the last flows. 
Moreover, we observe that starting the attack by the first flow may change the state of the target device. Thus, when the tool replicates the flow containing the actual command, the attack may fail because of the state change of the device and not because of an effective countermeasure implemented on the device.
\new{However, being our tool device-agnostic there is a lack of prior knowledge about which specific flow contains the actual command. Hence the need to replicate all flows rather than solely focusing on the last one.}

\subsection{Detection Module} \label{sec:detection_phase}
The aim of the detection module is to automatically detect whether the replay attack is successful.

We perform two preliminary checks: (i) response check and (ii) protocol check.

\noindent \textbf {Response Check:} If our tool receives no response during the attack phase, the attack is considered \textbf{FAILED}.

\noindent \textbf {Protocol Check:} If the companion app and the device communicate through standard security protocols (i.e., TLS, QUIC, etc.), the attack is considered \textbf{FAILED}.

In all the other cases, we assume a list of responses organized in a queue (the first response received during the attack phase is the first response scheduled in the detection phase).
We recall that, the flows during the attack phase are scheduled in reverse order with respect to the order of arrival.
We then expect that the response of an actual command (triggering a function) is contained in the first response stored by the queue.
In particular, as a heuristic approach, we consider the first $j$ responses received.

The ML algorithm receives as input 
these responses to check if they are detected as regular or irregular.
If \textbf{all} the $j$ responses are detected as irregular, we consider the attack \textbf{FAILED}.
Otherwise, we consider the attack \textbf{SUCCESSFUL}.
We assume that the fact that a device sends at least one regular response to a replayed request may be a security issue. This is the reason for which we consider the attack \textbf{SUCCESSFUL} when at least one response is detected as regular.

On the other hand, if in place of $j$, we feed all the responses sent by the target device to the ML algorithm, the chance of having false positives could be very high. To explain this, we consider an example. We suppose that, before sending a command, the companion app performs an initial handshake with the IoT device (e.g., exchanging keys). 
Then, by replaying the first request sent by the app (i.e., a request belonging to the initial handshake), the target device would most likely send a regular response. Then our tool would consider the attack as successful based on this response only, while it may not succeed in triggering the desired command. This is the reason why we limit the detection to the last $j$ responses.

\subsection{\tool Operation}

In this Section, we provide an example of the application of \tool. Figure~\ref{fig:proposed_framework} shows how \tool operates. We consider 6 steps, described below.

\noindent \textbf{Step 1:} We assume that a series of requests/responses are exchanged between the smartphone and the IoT device in the order shown in Step 1 of Figure~\ref{fig:proposed_framework}. 

\noindent \textbf{Step 2:} We sniff the traffic and organize it in flows. As an example, we consider that Flow 1 includes a single request ($A_1$) and a single response ($A_2$). Flow 2 includes a request ($B_1$) and two responses($B_2$ and $B_3$). Flow 3 includes two requests ($C_1$ and $C_2$) and a response ($C_3$).

\noindent \textbf{Step 3:} We perform the attack by scheduling the Flows in reverse order. 

\noindent \textbf{Step 4:} We replicate the captured requests to the device. In this example, we take into account that the device may not provide some responses  (e.g., because the state of the device is changed and the requests sent are not meaningful anymore).
Specifically, in this example, the response to the requests $C_1$ (or $C_2$) of Flow 3 is lost.
We observe that, in general, the received responses during the attack are not identical to the original responses (for example, they may contain different identifiers or timestamps). To highlight this, we use the overline notation ($\overline{C_3},\overline{B_2},\overline{B_3},\overline{A_2}$).

\noindent \textbf{Step 5:} We store the responses in the order we receive them for creating the queue.

\noindent \textbf{Step 6:} Finally, we use the responses to feed the detection module. 
 If the \textbf{Response Check} and the \textbf{Protocol Check} do not output \textbf{FAILED}, then the first $j$ responses are fed into the ML algorithm. In this example, we consider $j=2$. 


\section{Validation} \label{sec:validation}
\begin{figure}[t!]
	\centering
\includegraphics[width=1\columnwidth]{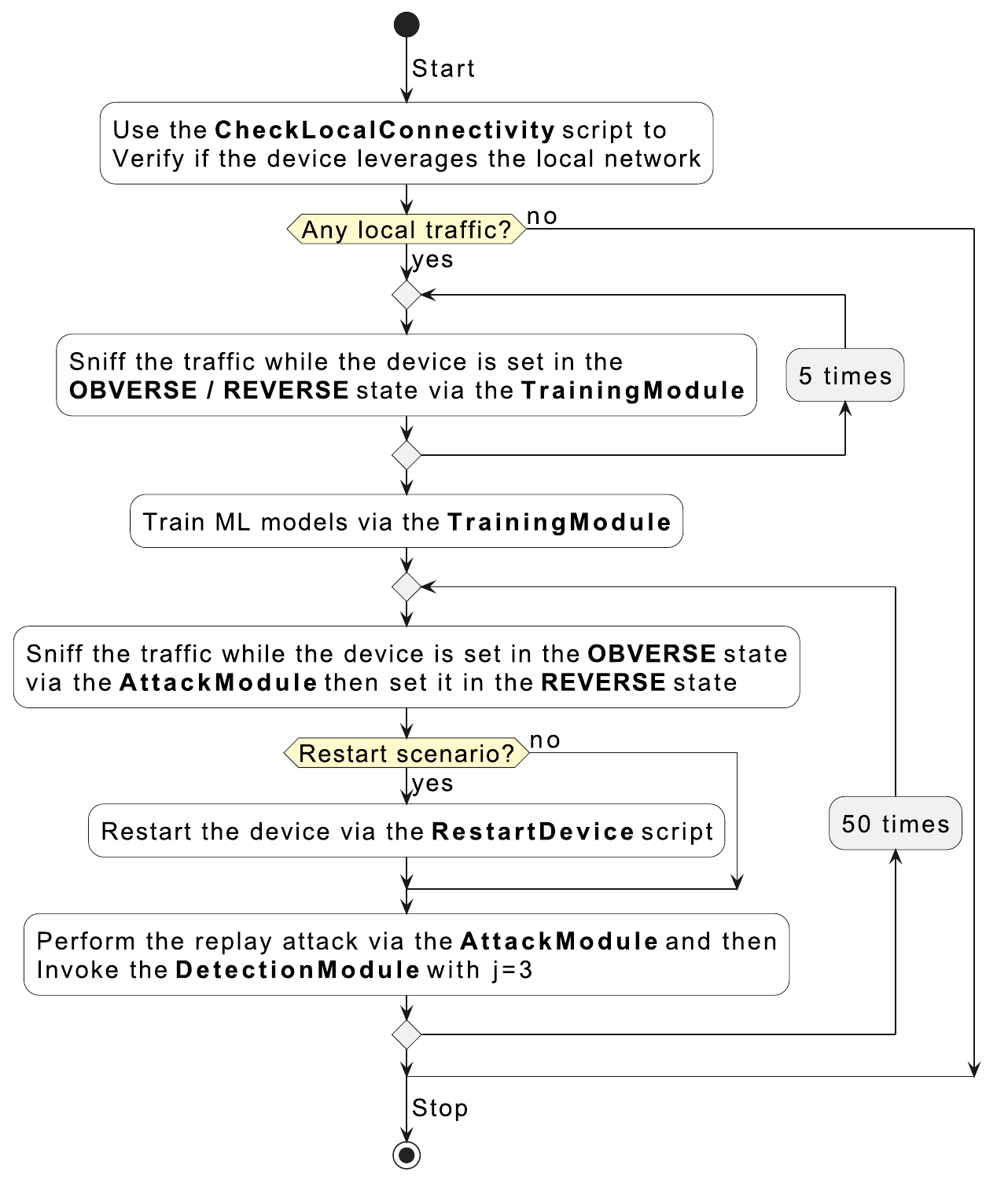}
     \caption{Procedure to Validate \tool.}
  \label{fig:tool_validation}
\end{figure}

In this section, we describe the procedure we follow to validate \tool and assess the effectiveness of 41 popular IoT devices in preventing replay attacks. 
Figure~\ref{fig:tool_validation} shows the steps of our validation.

\subsection{Testbed}
In order to have a controlled environment for testing the vulnerability to replay attacks of IoT devices, we build an IoT testbed.
Our testbed consists of: (1) 41 consumer IoT devices, (2) a smartphone in which we install the companion apps to control the IoT devices, (3) an access point (installed on a server) to which the smartphone and the target IoT devices are connected, (4) a set of support scripts to automatically turn on and off an IoT device, trigger a function, and determine whether the device has local connectivity.

\subsubsection{IoT Devices}
The devices we consider are consumer IoT devices typically deployed in a smart home. We have chosen devices under different categories (see the second column of Table~\ref{tbl:states}). \new{The devices are simultaneously connected to the network. We distinguish them by their MAC address.}

\subsubsection{Smartphone}
We use a standard smartphone (Google Pixel 3a) to manage the IoT devices through their apps.

\subsubsection{Access point}
The access point offers IP connectivity to the IoT devices under test, and the ability to capture and control the network traffic exchanged between the IoT devices and the apps. \tool is installed on the access point.

\subsubsection{Support Scripts}
We deploy the following 4 support scripts to automatize the validation procedure.
The {\tt CaptureCoordinates} script records the coordinates of the screen taps when we trigger a function to set the device in the \textbf{OBVERSE} or \textbf{REVERSE} state.
The {\tt ProbeCommand} takes such coordinates as input and uses them to automatically set the device in the \textbf{OBVERSE} or \textbf{REVERSE} state.
The {\tt CheckLocalConnectivity} script verifies whether the device and the companion app communicate through the local network.
Finally, the {\tt RestartDevice} script physically restarts a device by turning off a plug powering the device, and then turning it on.



In the following, we describe the steps followed to conduct the validation procedure.

\begin{table} [t!]
\caption{\textbf{OBVERSE} and \textbf{REVERSE} States per IoT Device}
\label{tbl:states}
\begin{tabular}{ p{3cm} |p{5cm} }
 \hline
\textbf{OBVERSE-REVERSE State} & \textbf{Device} \\ 
\hline

ON - OFF & Arlo pro 4 camera,  Govee lightstrip, Furbo camera, Lepro bulb, Lifx bulb, Meross smartplug, Nanoleaf triangle, Tapo smartplug, Wiz ligthbulb, Wyze cam pan, Yeelight bulb, Yeeligth lightstrip    \\ \hline

START - STOP & Coffee maker Lavazza, Cosori airfrier, Ecovacs vacuum, Eufy robovac, iRobot roomba i7, OKP vacuum, Swan Alexa Smart Kettle,      WeeKett Kettle, Xiaomi blender \\  \hline

PLAY - PAUSE & Blink camera, Blink video doorbell, Bose Speaker, Boyfun Baby monitor, Ring doorbell, Ring spotlight camera, Simplisafe camera,  Sonos Speaker, Vtech baby monitor   \\  \hline

ACTIVATE - DISACTIVATE & Aqara hub, Eufy doorbell, Meross Garage Opener, Switchbot hub mini  \\  \hline

MAX - MIN VOLUME & Echodot 4, Echodot 5, Eufy doorbell chime \\ \hline

 POSITIVE - NEGATIVE OFFSET & Netatmo Weather Station \\  \hline
SET KG-SET LB & Withings smart scale \\  \hline

SET °C-SET °F & Withings thermo  \\  \hline
 LOCK-UNLOCK & Petsafe Automatic Feeder \\  \hline

\end{tabular}

\end{table}

\subsection{State Phase}
This phase consists of finding the state in which the IoT device can be set in the companion app.
We assume that an IoT device cannot be in two states at the same time (e.g., the ON and OFF states).
We call one of these states as \textbf{OBVERSE} (i.e., smart light on) state and the other state as \textbf{REVERSE} (i.e., smart light off).

Table~\ref{tbl:states} shows the state \textbf{OBVERSE} and \textbf{REVERSE} considered for each IoT device in the testbed. 
For example, for the vacuum cleaners, the air fryer, the kettles, and the blender we consider the START-STOP state. The speakers and several cameras offer the PLAY-PAUSE state. For Withings smart scale and Withings thermometer, the states are represented by the unit of measurement for weight and temperature.

Since our experiments require to trigger the states several times, we adopt an automatic procedure for triggering them inspired by Mandalari et al.~\cite{mandalari2021blocking}'s study.
Specifically, we leverage the Android Debug Bridge (ADB) tool to programmatically trigger the function of the device by emulating the user's taps on the companion app. 
At first, we follow a one-time procedure. We manually perform some taps in the companion app to trigger the \textbf{OBVERSE} state and the \textbf{REVERSE} state.
Meanwhile, the {\tt CaptureCoordinates} script records the coordinates of the screen taps.
These coordinates are the input of the {\tt ProbeCommand} script. We then leverage this script to automatically set the device to the desired state.

To validate the correct execution of the {\tt ProbeCommand} script (setting the \textbf{OBVERSE} or \textbf{REVERSE} state on the device), we employ the following procedure.
Each time we invoke the {\tt ProbeCommand} script, we capture the screenshot of the companion app and compare it to a previously retrieved screenshot where we know that the command was correctly executed. 
For example, we invoke the {\tt ProbeCommand} script to turn on the smart light bulb and we capture the screenshot on the app to check whether the bulb is actually in state ON. 

\subsection{Check Local Connectivity} \label{sec:local_connectivity}
In this phase, we filter out the devices that do not communicate with the companion app in the local network but only via the cloud.
In general, some devices may work by default through the cloud even though the companion app and the IoT device are in the same network.
To verify this, we leverage the {\tt CheckLocalConnectivity}. This script forces the devices to communicate with their companion app via the local network (i.e., we prevent the devices from connecting to the Internet) by setting firewall rules at the access point.
This script starts using \textit{tcpdump} to capture the traffic in the local network.
Meanwhile, we invoke the {\tt ProbeCommand} script to set the device in the \textbf{OBVERSE} and the \textbf{REVERSE} state.
The {\tt CheckLocalConnectivity} script checks the presence of packets during this phase. If no packets are detected then the validation procedure ends (see Figure~\ref{fig:tool_validation}).

\subsection{Training Phase} 
\label{sec:training_validation}
This phase starts by invoking the {\tt Training Module}.
We recall that this module first launches {\tt tcpdump} to sniff the traffic sent from the device to the companion app when the latter issues some commands. We automatically trigger these commands via ADB through the {\tt ProbeCommand} script. Specifically, we first invoke this script to set the device in the \textbf{OBVERSE} state and then to set the device in the \textbf{REVERSE} state. We repeat this procedure five times.

Overall, the training set is composed of the responses to ten commands sent by the companion app (setting of \textbf{OBVERSE} and \textbf{REVERSE} states five times), along with possible control messages (e.g., synchronization messages or handshake messages).
\new{In the home environment, this procedure (i.e., the trigger of some commands) is performed manually by the users.}
We intentionally choose a small number of times (five) in which the two commands are repeated with the aim to easily get the dataset in the real-life application of the tool by users.
Finally, the {\tt Training Module} feeds the responses collected during the sniffing time to an ML model. 
For our validation, we select 3 standard models of ML for novelty detection~\cite{9004325,burlina2019s, PIMENTEL2014215}: Isolation Forest, Elliptic Envelope, and Local Outlier Factor.

\subsection{Attack Phase} \label{sec:val_attack}
During this phase, we first invoke the {\tt ProbeCommand} script to set the device in the \textbf{OBVERSE} state. 
The {\tt Attack Module} of \tool sniffs and processes the traffic exchanged by the device and the companion app (as described in Section~\ref{sec:attack_phase}).
We then invoke again the {\tt ProbeCommand} script to set the device in the \textbf{REVERSE} state. 

At this point, we perform the replay attack.
This attack is successful if \tool manages to switch the device from the \textbf{REVERSE} state to the \textbf{OBVERSE} state.
On the other hand, if the device remains in the \textbf{REVERSE} state, we assume the attack fails.
As ground truth, to check the final state of the device (\textbf{OBVERSE} or \textbf{REVERSE}), we perform the screenshot of the companion app screen and compare it with a previously captured screenshot taken when the device is known to be in the \textbf{OBVERSE} state.

We perform the above procedure in two different scenarios.

\begin{itemize}
    \item \textbf{Non-Restart Scenario}: in this scenario, we first set the device in the \textbf{REVERSE} state and then we perform the replay attack. 
    
    \item \textbf{Restart Scenario}: in this scenario, we first restart the device via the {\tt RestartDevice} script, and then we set the device in the \textbf{REVERSE} state. Finally, we perform the attack.
\end{itemize}

The two scenarios capture two different situations. The \textbf{Non-Restart Scenario} 
represents a real-life scenario, in which we simulate a replay attack after a user legitimately set the device in the \textbf{REVERSE} state. In this case, the effect of the replay attack is to change the state of the device against the will of the user.
We point out that this is not the simplest scenario in which we test \tool. Indeed, a \textit{Real-time} attack, in which we replicate a command immediately after capturing it, may have more chances to be successful.
However, the impact of such an attack would be minimal since the device would be set in the same state desired by the user. 
On the other hand, the \textbf{Restart Scenario} is a more challenging scenario in which to test our tool. 
Indeed, the restart forces the device to clear its working memory (possibly negotiated temporary keys, tokens, agreed ports, etc.). When the \textbf{REVERSE} state is set, there are more chances that the device would possibly renegotiate some values (i.e., keys, tokens, ports, and so on), thus making the attack ineffective.
In our validation, we perform the attack 30 seconds after the restart to give the device the time to run the initial configuration procedure.

\subsection{Detection Phase}
In this phase, we invoke the {\tt Detection Module} of \tool. We recall that this module takes as input the packets (properly pre-processed) sniffed by the {\tt Attack Module}. Moreover, this module requires a parameter $j$ to be set representing the number of responses to provide to the ML algorithm. In our validation procedure, we set $j=3$.
\new{This value is set empirically by testing manually a subset of devices.}

To evaluate the performances of the {\tt Detection Module}, we invoke the {\tt Attack Module} 50 times. For each of these, we invoke the {\tt Detection Module} that returns a boolean output: whether the replay attack is successful or not. 
We observe that the ground truth remains the same for all 50 experiments, i.e., the attack either always works or does not.
To measure the {\tt Detection Module} performance, for a given device, we consider the \textit{accuracy} measured as the number of times in which the {\tt Detection Module} successfully classifies the attack over the total number of experiments.
The \textit{accuracy} coincides with the \textit{recall} while the \textit{precision} is not significant because it is always $1$. Indeed, by considering as positive instance the result of the ground truth, there are no false positive and true negative instances.
Then, the \textit{accuracy} (i.e., the recall) is the only meaningful metric to consider.

\subsection{Validation through API} \label{sec:api}
In case no local traffic is exchanged or the replay attack is not successful, we consider an additional step. 
\new{We check whether the IoT device maintains some open ports for developing personal applications and communicates with the device through some APIs \cite{localAPI}. This interaction is referred to as local APIs.
}
The main goal of our investigation remains the interaction between the companion app and the target device. 

To test the robustness of devices when APIs are adopted, we implement ad hoc scripts for these devices, emulating the behavior of the smartphone through APIs. Then, we repeat the methodology described in Figure~\ref{fig:tool_validation}, with the difference that, this time, the device is not set in the \textbf{OBVERSE/REVERSE} state through the {\tt ProbeCommand} script but through customized scripts using the available APIs.

\section{Results} \label{sec:result_discussion}
We answer our research questions by applying the methodology described in Section~\ref{sec:methodology} and \ref{sec:validation} to understand whether it is possible to automate replay attacks against IoT devices and automatically detect whether the attack is successful. 

\subsection{Local Connectivity} \label{subsec:localConnectivity_resul}

By leveraging the procedure described in Section~\ref{sec:local_connectivity}, we find that out of the 41 devices of our testbed, 22 do not communicate with the related app through the local network. Among the devices that do not use local connectivity, one (the Govee lightstrip) uses local APIs.
This means that 21 devices out of 41 are not compliant with the principle stated in the ENISA security guideline (GP-TM-17~\cite{ENISA-IoT}), stating that an IoT device should continue to function if the cloud back-end fails.

\noindent\textbf{Takeaways.}
The finding that, among 41 devices, 21 lack local connectivity, is inherently significant. It underscores the pressing need for substantial improvements in the reliability of IoT devices. \new{In addition, allowing devices to work independently of the cloud enables users to have greater control over their data and reduces the risk of data exposure during cloud outages or breaches. Finally, local connectivity reduces the dependency on the cloud, thus resulting in lower latency and energy saving.}

\subsection{Replay Attack} \label{subsec:result_replayAttack}

Table~\ref{tbl:replayResult} shows the result of the performed replay attacks on the 20 devices that leverage local connectivity in the two considered scenarios (i.e., \textbf{Non-Restart} and \textbf{Restart}).
\new{To fairly validate our results, besides using the screenshot procedure as described in Section \ref{sec:val_attack}, we manually verify whether the replay attack is working or not for each device.}

\begin{table}[t]
\caption{Replay Attack Results. $\cmark$ indicates whether the replay attack is successful or not ($\xmark$).}
\label{tbl:replayResult}

\begin{tabular}{ p{3.5cm} |p{1.75cm} |p{1.75cm} }
 \hline
\textbf{Device ($^*$Tested via APIs)} & \textbf{Non-Restart Scenario}  & \textbf{Restart Scenario} \\ 
\hline

Bose Speaker $^*$ & $\cmark$ & $\cmark$ \\  \hline

Boyfun Baby monitor & $\xmark$ & $\xmark$ \\  \hline

Eufy robovac  & $\cmark$ & $\cmark$ \\  \hline

Furbo camera & $\xmark$ & $\xmark$ \\  \hline

Govee lightstrip $^*$ & $\cmark$ & $\cmark$ \\  \hline

iRobot roomba i7 & $\xmark$ & $\xmark$ \\  \hline

Lepro bulb  & $\cmark$ & $\cmark$ \\  \hline

Lifx bulb  & $\cmark$ & $\cmark$ \\  \hline

Meross smartplug  & $\cmark$ & $\cmark$ \\  \hline

Meross Garage Opener & $\cmark$ & $\cmark$ \\  \hline

Nanoleaf triangle $^*$ & $\cmark$ & $\cmark$ \\  \hline

OKP vacuum & $\cmark$ & $\cmark$ \\  \hline

Sonos Speaker  $^*$ & $\cmark$ & $\cmark$ \\  \hline

Tapo smartplug  & $\cmark$ & $\xmark$ \\  \hline

Vtech baby monitor &  $\xmark$ & $\xmark$ \\  \hline

WeeKett Kettle  & $\cmark$ & $\cmark$ \\  \hline

Wiz ligthbulb & $\cmark$ & $\cmark$ \\  \hline

Wyze cam pan & $\xmark$ & $\xmark$ \\  \hline

Yeeligth lightstrip & $\cmark$ & $\cmark$ \\  \hline

Yeelight bulb & $\cmark$ & $\cmark$ \\  \hline
\end{tabular}

\end{table}

As reported in Section~\ref{sec:api}, in case no local traffic is exchanged or the replay attack is not successful, we test whether IoT devices present vulnerable local APIs. This happens for 4 devices (i.e., Govee lightstrip, Bose speaker, Sonos speaker, and Nanoleaf triangle). We denote these devices by $^*$ in Table~\ref{tbl:replayResult}. 
Specifically, as reported in Section~\ref{subsec:localConnectivity_resul}, the Govee lightstrip does not leverage the local network when communicating with the companion app.
On the other hand, the Bose speaker and the Sonos speaker leverage TLS to communicate with their companion apps. Then, our tool is not able to perform the attack.
Finally, the Nanoleaf triangle uses a secure proprietary protocol based on HTTP to communicate with its companion app, and again the attack is not successful. 
Despite this, we found that these devices present local APIs vulnerable to replay attacks.

Surprisingly, among the 20 devices, 15 are found to be vulnerable to the replay attack. This attack successfully exploits all the devices in the \textbf{Non-Restart Scenario} and is effective against 14 of them in the \textbf{Restart Scenario}.

To investigate the reasons behind our results, we manually study the response payloads of the 20 devices during the replay attack.
As reported in Section~\ref{sec:observed_responses}, we observe four types of responses in the local traffic: \textbf{full cleartext}, \textbf{standard encrypted}, \textbf{non-standard encrypted}, \textbf{encoded}.
Table~\ref{tbl:response_distribution} describes the type of responses adopted by each device. 
We observe that 9 devices adopt cleartext responses, 5 devices use encoded responses, 5 devices adopt non-standard encrypted responses, and only 1 device (i.e., the iRobot roomba i7) employ standard encrypted responses (i.e., TLS).

\begin{table}[t]
\caption{Response Distribution}
\label{tbl:response_distribution}
\begin{tabular}{ p{2.5cm} |p{5.5cm}}
 \hline
\textbf{Response} & \textbf{Device}  \\
\hline

\textbf{Cleartext} & Bose Speaker,
 Govee lightstrip,  Meross Garage Opener, Meross smartplug, Nanoleaf triangle, Sonos Speaker,  Wiz ligthbulb, Yeelight bulb, Yeeligth lightstrip     \\  \hline

\textbf{Standard Encrypted} & iRobot roomba i7   \\ \hline

\textbf{Non-Standard Encrypted} & Boyfun Baby monitor, Furbo camera,   Tapo smartplug, Vtech baby monitor, Wyze cam pan  \\ \hline

\textbf{Encoded} & Eufy robovac, Lepro bulb, Lifx bulb,  OKP vacuum, WeeKett Kettle \\ \hline

\hline

\end{tabular}
\end{table}

\begin{table}[t]
\caption{Examples of Responses' Payload}

\label{tbl:payloads}
\begin{tabular}{ p{1.9cm} |p{6.2cm}}
 \hline
\textbf{Device} & \textbf{Payload}  \\
\hline

\textbf{WeeKett Kettle} &  ËR<ñüg©§EDlyÿ:CÆFÎ¦íyB6PUª°QªU\\ \hline

\textbf{Tapo smartplug} &
\makecell{\{\{``error\_code":0,``result":\{``response":``CO07WBT2x\\BhRL05oIZbhAEuf/FjQEEa596JE3+X1ubE="\}\}}  \\  \hline

\textbf{Meross Garage opener} & \makecell{``header":\{``messageId":``08..6c",``namespace":``Ap-\\pliance.System.DNDMode",``triggerSrc":``Android-\\Local",``method":``SETACK",``payloadVersion":1,\\``from":``/appliance/22..1b/publish",``timestamp":16..90,\\``timestampMs":814,``{\color{red}sign}":``41..3d"\},``payload":\{\}\}} \\ \hline

\end{tabular}
\end{table}

Table~\ref{tbl:payloads} describes response payloads interesting to discuss.

In the \textbf{Restart Scenario}, the 14 devices vulnerable to replay attacks all communicate through protocols that adopt either encoded or cleartext responses. 
In Table \ref{tbl:payloads}, we give an example of an encoded response for a smart kettle. This response (and also the associated request) always presents the same payload.
These devices do not perform any authentication with the companion app after rebooting.

The Tapo smartplug is vulnerable only in the \textbf{Non-Restart Scenario}. This device leverages a  proprietary protocol that includes partially encrypted responses (see Table~\ref{tbl:payloads}). However, this protocol is not secure since an encrypted packet remains valid over time until the device is rebooted (\textbf{Restart Scenario}). In this case, the device exchanges a new key with the companion app and thus the attack is not successful. 
We point out that the \textbf{Non-Restart Scenario}  is the most plausible in real-world situations. 

Two devices that deserve attention are the Meross smartplug and the Meross Garage opener.
These devices leverage cleartext messages that include a signature to prevent commands from being altered. However, the signature remains valid when the tool replicates the same message so that the devices are vulnerable.
We report in Table~\ref{tbl:payloads} an example of a response that includes such a signature (highlighted in red).

We now focus on the remaining 5 devices in which the replay attack is not successful.
Four of the devices are smart cameras.
A manual inspection reveals that the cameras examined use (on the local network) proprietary protocols that do not appear to be vulnerable to \tool.
We observe that this does not guarantee that they are not vulnerable to replay attacks as they do not adopt standard security protocols. As such, our
findings represent a lower-bound of such vulnerability, using
an approach that can be automated.
The iRobot roomba i7 communicates with the companion app via the TLS protocol. Generally speaking, the adoption of standard security protocols should be preferable to counter replay attacks.

\noindent\textbf{Takeaways.}
Our findings demonstrate the effectiveness of \tool. 
Despite its agnostic nature, \tool can automatically identify 15 out of 20 devices vulnerable to replay attacks, thus positively answering the research questions \textbf{RQ1} and \textbf{RQ3}.
In addition, our manual investigation highlights that IoT devices use proprietary protocols with weak (or not) security measures. This makes them vulnerable to replay attacks.
\new{Finally, out of 20, 14 devices are still vulnerable to replay attacks in the \textbf{Restart Scenario}. This denotes a lack of an authentication procedure. To prevent the attack from working, well-known security protocols with mutual authentication mechanisms (e.g., TLS with embedded certificates) should be adopted. }

\subsection{Detection Module} \label{sec:detection_accuracy}
\new{As reported in Section~\ref{sec:training_validation}, we investigate three different ML algorithms. We find that the Local Outlier Factor algorithm outperforms the other two algorithms.
Specifically, with the Local Outlier Factor model, the accuracy of the {\tt Detection Module} ranges (over all the devices) from 0.98 to 1 for both the \textbf{Restart} and \textbf{Non-Restart} scenarios.
Concerning the Isolation Forest model, we have an accuracy of 0\% for a single device, while for the other devices the accuracy ranges from 0.98 to 1. Regarding the Elliptic Envelope model, we have an accuracy of 0\% for two devices, while for the other devices the accuracy ranges from 0.78 to 1.
}


We recall that the {\tt Detection Module} performs two preliminary checks, before (possibly) feeding the responses to the ML algorithm.
The \textbf{Response Check} consists of verifying if the device under test responds to our tool during the attack phase. It outputs \textbf{FAILED} when our tool receives no responses.
The \textbf{Protocol Check} consists of verifying if the device adopts standard security protocols for local communication. If so, it outputs \textbf{FAILED}.
If none of the two checks output \textbf{FAILED}. The {\tt Detection Module} feeds the received responses to the ML algorithm.


Table~\ref{tbl:detection_moduleIntervention} shows how the {\tt Detection Module} works for each device. 
Specifically, we observe that for 4 devices the \textbf{Response Check} outputs FAILED and for 1 device the \textbf{Protocol Check} outputs FAILED.

\begin{table}[t]
\caption{{\tt DetectionModule} Results}
\label{tbl:detection_moduleIntervention}
\begin{tabular}{ p{3cm} |p{5cm}}
 \hline
\textbf{Detection} & \textbf{Device}  \\
\hline

\textbf{Response check} & Boyfun Baby monitor, Furbo camera, Vtech baby monitor,  Wyze cam pan   \\  \hline

\textbf{Protocol check} &  iRobot roomba i7   \\ \hline

\textbf{ML intervention} & Bose Speaker, Eufy robovac, Govee lightstrip, Lepro bulb, Lifx bulb, Meross Garage Opener, Meross smartplug,  Nanoleaf triangle, OKP vacuum, Sonos Speaker, Tapo smartplug, WeeKett Kettle, Wiz ligthbulb, Yeelight bulb, Yeeligth lightstrip\\ \hline

\end{tabular}
\end{table}

There are two main reasons to explain the high accuracy values of the {\tt Detection Module}.
First, the two preliminary checks (avoiding the intervention of the ML algorithm) allow us to effectively detect that the replay attack does not work. This is the case for 5 devices where either our tool receives no response or the devices use secure communication protocols.
Secondly, when the ML algorithm is invoked (for the remaining 15 devices), the responses fed to it have many similarities.
Indeed, by manual investigation, we observe that for some devices the responses are exactly the same while for other devices just a few fields (e.g., identifiers and timestamps) change. Hence, the ML algorithm readily identifies this similarity.

\noindent\textbf{Takeaways}
The performance achieved by the {\tt Detection Module}, included in \tool, shows an accuracy value ranging from 0.98 to 1, thus positively answering the research question \textbf{RQ2}.
\new{Accuracy is crucial for determining the efficacy of our tool for autonomously detecting vulnerabilities to replay attacks in IoT devices. This is fundamental since the tool's intended adoption is by non-technical users in domestic settings. The detection module is particularly relevant in the case in which the user launching the tool has no physical access to the device to observe the effect of a replayed command, or the effect of the command may not trigger any visible changes in the device status.}

The results obtained encourage us to believe that our tool can be adopted (beyond the lab) by non-technical users to detect whether the IoT devices they own are vulnerable to replay attacks.

\section{Discussion}
\label{sec:discussion}
In this section, we discuss the implications of our findings, possible mitigation, limitations, and ethical considerations.

\noindent \textbf{Safety and Reliability implications.}
We highlight that the absence of local connectivity (detected in more than 50\% of our devices) may harm the safety and reliability of an IoT device. This principle is stated in the ENISA security guideline GP-TM-17~\cite{ENISA-IoT}, stating  \textit{``essential features should continue to work with a loss of communications and chronicle negative impacts from cloud-based systems ... a loss of communications shall not compromise the integrity of the device, and the device should continue to function if the cloud back-end fails''}.

Safety problems arise also when local connectivity is supported but replay attacks are effective.
For example, among vulnerable devices, we found two different smart plugs, a garage opener, \new{and a smart kettle}.
The impact of the attack on these kinds of devices is particularly critical in terms of safety. Indeed a smart plug can be used to power a generic device (it may be a medical device or a security camera). Thus a vulnerable plug may directly affect the user safety.
\new{Similarly, a replayed ``open'' command sent to a garage door opener may enable a domestic violation.
Also, if a heat-related device (such as a smart kettle) is turned on without any authorization, there might be a risk of overheating, thus potentially leading to a fire. This could 
pose a threat to the safety of the home occupants.
It is also worth noting that unauthorized activation of a generic smart device could lead to unplanned and excessive energy consumption.}

\noindent \textbf{Security and Privacy implications.}
The effectiveness of our tool denotes a lack of proper security measures on the IoT devices in the local network. This violates several guidelines of ENISA and NIST (see Section~\ref{sec:assumptions}).
Our results clearly highlight that devices rely on the security provided by firewalls and access points, rather than offering security mechanisms themselves (in contrast with GP-TM-43 of ENISA).

While the impact in terms of security is evident (devices can be managed without users' will), we observe that replay attacks may lead to privacy issues.
For example, the activation of a recording function on a smart speaker may be used to record the voice of a user without their explicit consent, opening the door to new potential IoT tech-abuse phenomena~\cite{parkin2019usability}.

\noindent \textbf{Mitigation.}
Possible mitigations could include the adoption of a mutual authentication procedure. This would ensure that only authorized parties can issue commands to the IoT device.
This solution should always be adopted in conjunction with standard security protocols. Indeed, we can find an attempt at an authentication procedure in the Tapo smartplug. Unfortunately, it is not performed via a security protocol, as it suffers from a lack of randomness and key reuse. Hence this device is still vulnerable to replay attacks.
\new{We suggest the adoption of WPA3 that prevents traffic interception from a malicious device in the same network also in the case the network password is known. However, it is not effective in the case of a compromised access point. In addition, we point out that WPA2 is still widely deployed, thus the threat we identified is very likely.
}

\noindent \textbf{Limitations.}
While we have made our best effort to investigate how IoT devices react to reply attacks, as a first attempt at this space, this work has a few limitations.

\noindent \emph{Assessing Device Vulnerability.}
The agnostic nature of our tool does not allow us to leverage customized vulnerabilities of a target device. As a consequence, when our tool fails to perform a replay attack on the device, it is not a guarantee that the device is not vulnerable. As such, our
findings represent a lower bound of such vulnerability, using an approach that can be automated, i.e., automatically detecting the effectiveness of the attack.
On the other hand, the fact that out of 20 potentially vulnerable devices, \tool succeeds in performing the attack against 15 devices, confirms the efficacy of our tool. 

\noindent \emph{Scalability.} 
While every step of the validation of our approach is fully automated, including the execution of function trigger scripts, the creation of these scripts is a manual process that needs to be repeated individually for every function tested on each device. One mitigating factor is that devices within the same categories can often reuse existing scripts with minimal modifications. 
Lastly, our study is limited to the number of popular devices in our testbed, and we do not investigate all IoT devices on the market. However, \tool is designed to be device-agnostic and will easily scale to other devices. Testing more devices and performing longitudinal studies will be a valuable next step in future works; \new{hence we make publicly available our tool and data, to ease the reproducibility of our experiments.}

\noindent \emph{Non-Observable Functions.} 
Our approach is designed to operate exclusively with device functions that can be assessed using trigger scripts. Certain functions, such as device maintenance or synchronization tasks, cannot be initiated directly. To test whether the attack works with these functions, one can follow manual steps.

\noindent \textbf{Ethical Considerations.} 
In our experiments we do not cause
any real threat on the Internet. All experiments are contained
within our own testbed. No traffic related to human activity was collected during the experiments.  When conducting the experiments, we fully respected the ethical guidelines defined by our affiliated organization, and we received approval. In testing the functionality of \tool, we exclusively test our own devices, eliminating the privacy
risk to others.

\noindent
\textbf{Feedback from Vendors.}
To this date, only TP-Link, the manufacturer of the Tapo smart plug, has acknowledged the identified issue and taken proactive steps by releasing a new firmware to address the vulnerability. This corrective action extends its positive impact to millions of Tapo devices, underscoring the significance of our work in enhancing the security and resilience of IoT devices.

\section{Related Work} \label{sec:related_work}

In this section, we discuss the scientific literature related to replay attacks in the IoT domain. To the best of our knowledge, no proposal in the literature pursues the same goals of this paper. 
The problem of replay attacks in IoT is acknowledged in the literature~\cite{lu2018internet}. As a consequence, several scientific works~\cite{feng2017replay, de2020energy, farha2020timestamp} propose solutions to mitigate it. However, no commercial device currently implements them.

Other papers are devoted to the detection of replay attacks against IoT devices~\cite{elsaeidy2020replay, elsaeidy2021hybrid, 9996544}. 
Similarly,~\cite{malik2020light,malik2019towards,pradhan2019combating} focus on detecting replay attacks against voice-activated services. However, the primary purpose of these works is to identify whether a device is currently under a replay attack and these solutions are integrated as part of an intrusion detection module, thus proposing mitigation techniques for replay attacks. In contrast, as explained in Section~\ref{non-goals} (Non-Goals), our paper addresses the active automated execution of replay attacks with the goal of assessing the vulnerability of devices to replay attacks.

Wara and You~\cite{wara2020new}'s study
is the closest to our work since it shows how replay attacks can be performed on IoT devices supporting ZigBee. However, the tool proposed is not agnostic and the experimental evaluation is quite limited (the experimental validation is conducted on three devices only).

Other works~\cite{8767399,yoon2020new} actively perform replay attacks on smart speakers. However, again, they are not device-agnostic and the replay attack refers to record and replay the voice of users to trigger commands, not testing generic functions of a consumer IoT device.

\new{Finally, our work is closely aligned with a relevant technique known as IoT fuzzing~\cite{eceiza2021fuzzing}. However, there are some inherent distinctions.
Fuzzing is not primarily designed for detecting or preventing replay attacks. Fuzzing is a software testing technique that involves sending intentionally malformed or random data to a target to discover vulnerabilities. As a matter of fact, our tool aims to replay legitimate messages to verify whether possible authentication mechanisms put in place by an IoT device can be bypassed. Indeed, our tool preserves the original messages including potential signatures, authentication tokens, nonces etc., thus it has more chances than a fuzzer to craft messages that will be accepted by IoT devices.}

\section{Conclusion} \label{sec:conclusion}

Due to the proliferation of IoT devices in smart home, the protection of the local network segment is a crucial aspect to take into consideration.
In this paper, we focused on replay attacks that can be performed by an adversary with access to the local network to trigger functions on the devices without the user's will.
We developed a tool, called \tool, for automatically testing replay attacks on consumer IoT devices.
The tool is designed to be device-agnostic, thus not requiring prior knowledge of the specific devices.
We employed this tool to perform a large-scale experiment involving 41 devices spanning different vendors and categories.

Our experiments reveal the existence of several vulnerable devices or not compliant with safety guidelines, thus demonstrating that this threat is real and can potentially harm users' households.

\new{As future work, we plan to extend our tool with new features. Specifically, we intend to leverage Natural Language Processing (NLP) techniques, enabling us to differentiate the diverse commands associated with devices, facilitating the precise triggering of specific actions.} 

\new{To support further research, all software and data
we produced as part of this work are publicly available
at \url{https://github.com/SafeNetIoT/ReplayAttack}.} 

\section*{Acknowledgements}
\new{
We thank the anonymous reviewers and our shepherd Urs Hengartner for their constructive feedback. This work is partially funded by project SERICS (PE00000014) under the MUR National Recovery and Resilience Plan funded by the European Union - NextGenerationEU, and the EPSRC PETRAS (EP/S035362/1).}


\newpage
\balance


\end{document}